\documentclass{aa}
\usepackage{epsfig}
\input{epsf}
\def\lsim{\lower.4ex\hbox{$\;\buildrel <\over{\scriptstyle\sim}\;$}}
\begin{document}
\title{Differential rotation decay in the radiative envelopes of CP stars}
\titlerunning{Differential rotation decay in CP stars}
\author{Rainer Arlt\inst{1} \and Rainer Hollerbach\inst{2} \and G\"unther R\"udiger\inst{1}}
\authorrunning{R. Arlt, R. Hollerbach, G. R\"udiger}
\institute{Astrophysikalisches Institut Potsdam, An der Sternwarte 16, 
D-14482 Potsdam, Germany\and Brandenburgische Technische Univ.\ Cottbus, 
Siemens-Halske-Ring 1, D-03044 Cottbus, Germany}

\date{\today}

\abstract{Stars of spectral classes A and late B are almost entirely radiative.
CP stars are a slowly rotating subgroup of these stars. It is
possible that they possessed long-lived accretion disks in their
T Tauri phase. Magnetic coupling of disk and star leads to rotational
braking at the surface of the star. Microscopic viscosities
are extremely small and will not be able to reduce the rotation
rate of the core of the star. We investigate the question
whether magneto-rotational instability can provide turbulent
angular momentum transport. We illuminate the question whether or not
differential rotation is present in CP stars. Numerical MHD simulations of 
thick stellar shells are performed. An initial differential rotation 
law is subject to the influence of a magnetic field. The configuration
gives indeed rise to magneto-rotational instability. The emerging
flows and magnetic fields transport efficiently angular momentum 
outwards. Weak dependence on the magnetic Prandtl number ($\sim10^{-2}$
in stars) is found from the simulations. Since the estimated time-scale 
of decay of differential rotation is $10^7$--$10^8$~yr and comparable 
to the life-time of A stars, we find the braking of the core to be an 
ongoing process in many CP stars. The evolution of the surface rotation 
of CP stars with age will be an observational challenge and of much
value for verifying the simulations.
\keywords{stars: chemically peculiar -- stars: rotation -- 
stars: magnetic fields -- MHD -- turbulence}
}

\maketitle

\section{Introduction}
The present work was originally motivated by the existence
of stars of spectral type A of which a subgroup is peculiar
to their strong magnetic fields. These peculiar A and late B 
stars, which are collectively called CP stars, rotate 
significantly slower than their non-magnetic relatives.
The dipole axes of the magnetic fields of CP stars have various
orientations, with a tendency to larger tilts at faster 
rotation (Landstreet \& Mathys 2000). 

Young stars suffer from considerable rotational braking 
in particular during their pre-main sequence evolutionary
phase. Angular momentum is lost partly through stellar winds
in the T Tauri phase. Additional braking probably applies to 
T Tauri stars having an accretion disk (classical T Tauri 
stars -- CTTS). These stars also form a slowly rotating 
subgroup among the T Tauri stars (Bouvier et al.\ 1993) just 
as the slow CP stars among ordinary A and late B stars. Magnetic 
fields exiting the accretion disk will couple to the stellar field and
exert torques as well. The braking is efficient for solar-mass 
T Tauri stars as found by Cameron \& Campbell (1993). Recently,
St\c{e}pie\'n (2000) proposed the analogy between the CP
star subgroup and the CTTS among T Tauri stars and explains
the slow CP stars by rotational locking with the disk in
their pre-main-sequence phase.

If stars are considerably slowed down at the surface,
how fast rotates their interior? The microscopic viscosity
of stellar plasma is extremely small. We can estimate the 
time-scale of viscous decrease of the rotation
in case of external braking simply by estimating the time-scale
due to the gas' viscosity of roughly $\nu=10$~cm$^2$/s.
For a stellar radius of $R=10^{11}$~cm, we get a viscous time-scale
of $R^2/\nu \approx 10^{13}$~yr which is 4--5 orders of magnitudes
longer than the life-time of A stars and even a thousand times 
more than the age of the universe. Viscous decay of
the differential rotation in any radiative shell of a star
is thus not applicable. As long as there is no convection
providing turbulent transport of angular momentum
(whose efficiency may, however, not be very high),
we should expect strong differential rotation on the
way from the stellar surface to the deep interior.

A differential rotation in a radiative star will be prone
to the magneto-rotational instability which requires only
two things: an angular velocity decreasing with axis distance
and a weak magnetic field (Balbus \& Hawley 1991). The question 
what ``weak'' means will be addressed in Section~2 in the discussion 
of the initial magnetic field used for our simulations.
The instability is known to evolve quickly on the time-scale of
the rotation period. It has turned out to be an efficient generator 
of turbulence in accretion disks. See e.g.\ Kitchatinov \& R\"udiger 
(1997) for a linear global analysis and Hawley (2000) and Arlt 
\& R\"udiger (2001) for simulations. However, the instability 
is as well applicable to a stellar interior as long as the 
angular velocity decreases with axis distance in parts of 
the spherical domain. The mechanism is typically termed 
magneto-rotational instability or Balbus-Hawley instability. 
It is a consequence of the local and linear MHD equations; 
a lower limit to the magnetic field is only imposed by the 
magnetic diffusivity $\eta$ which is extremely small for 
stellar plasma. Field geometry is also almost irrelevant 
for the onset of the instability. The magneto-rotational 
instability must be quite ubiquitous in stellar radiative 
zones as soon as differential rotation emerges, likely to be
caused by surface braking.

We would like to address the question of 
how long it takes to turn the radiative envelope of the star into
a uniformly rotating shell. Is there any chance to maintain
a rotation profile depending only on the axis distance?
The answer on whether or not differential rotation may
be expected, will have consequences for the geometry of
magnetic fields or possibly for their generation in CP stars.

The questions were investigated numerically. In the second 
Section we will describe our setup for the simulations. The
third and fourth Sections deal with the hydrodynamic and
magnetohydrodynamic evolutions of a thick spherical shell.
We will discuss the consequences of the computations in
Section~5.

\section{Simulations\label{setup}}
The simulations apply the spectral, spherical MHD code of Hollerbach
(2000). The computational domain covers a full spherical shell
from the inner radius $r_{\rm i}=0.2$ to the outer 
radius $r_{\rm o}=1.0$. We start from the non-ideal
MHD equations with the kinematic viscosity $\nu$ and the magnetic 
diffusivity $\eta$. The time-dependent, incompressible, non-dimensional 
equations are then
\begin{eqnarray}
\Bigl(\frac{\partial}{\partial t} + \vec{u}\cdot\nabla\Bigr)\vec{u}&=& 
  -\nabla p
  + {\rm Pm} \nabla^2\vec{u}
  + (\nabla\times \vec{B})\times \vec{B}\label{ns}\\
\frac{\partial \vec{B}}{\partial t} &=& \nabla^2 \vec{B} 
  + \nabla\times(\vec{u}\times \vec{B})\label{induction}
\end{eqnarray}
with the usual meanings of $\vec{u}$, $\vec{B}$, and $p$ as the velocity, 
magnetic field, and pressure which is not explicitly
calculated in this model, but eliminated by applying
the curl-operator to Eq.~(\ref{ns}). Lengths are normalized with 
the radius of the sphere, $R$, times are measured in diffusion times 
$R^2/\eta$, velocities are normalized with $\eta/R$ as well as 
magnetic fields with $\sqrt{\mu\rho} \eta/R$. Note that the 
permeability $\mu$ and the density $\rho$ are constants in 
our approach. This normalization leads to the magnetic Prandtl number,
\begin{equation}
  {\rm Pm} = \frac{\nu}{\eta},
\end{equation}
measuring the ratio of diffusive to viscous time-scales.

Instead of the physical $\vec{u}$ and $\vec{B}$, we integrate the
potentials $e$, $f$, $g$, and $h$ which compose the physical
quantities by
\begin{eqnarray}
  \vec{u} &=& \nabla\times(e\vec{\hat r}) + \nabla\times\nabla\times(f\vec{\hat r}),\\
  \vec{B} &=& \nabla\times(g\vec{\hat r}) + \nabla\times\nabla\times(h\vec{\hat r}),
\end{eqnarray}
where $\vec{\hat r}$ is the unit vector in radial direction. 
The potentials for $\vec{u}$ and $\vec B$ are decomposed into
Chebyshev polynomials for the radial dependence and into
spherical harmonics for the angular dependence.
The representation by potentials implies that $\nabla\cdot\vec u=0$ and
$\nabla\cdot\vec B=0$ are always fulfilled automatically.

The initial conditions for the velocity represents a rotation
profile in which the angular velocity decreases with the 
cylinder radius, $s=r\sin\theta$, according to
\begin{equation}
  \Omega = \frac{{\rm Rm}}{\sqrt{1 + (2 s)^{2q}}},
  \label{omega}
\end{equation}
where Rm is the magnetic Reynolds number which is determined by the
normalized angular velocity on the axis,
\begin{equation}
  {\rm Rm} = \frac{R^2 \Omega_0}{\eta}.
\end{equation}
This Rm will be varied in our simulations; we always put $q=2$. 
A profile depending on the axis distance appears to be a reasonable
choice for the internal rotation of a star being prone to magnetic
coupling with an accretion disk. 

According to the Rayleigh criterion of hydrodynamic stability,
\begin{equation}
  \frac{dj^2}{ds}>0,
\end{equation}
where $j=s^2\Omega$ is the angular momentum per unit mass, the
above rotation profile (\ref{omega}) will provide us with a
hydrodynamically stable configuration. For large axis distances $s\gg1$
the profile with $q=2$ would be marginally stable, but within
our shell of finite radius, the Rayleigh criterion is not
violated. 

The construction of the initial magnetic field is based on a
vertical, homogeneous field, onto which we impose a non-axisymmetric
perturbation of Fourier mode $m=1$. The total initial magnetic field
can be written as
\begin{equation}
\vec{B} = B_0 [\vec{\hat z} + \epsilon \sin (kz + \pi/4) \vec{\hat x}],
\label{perturbation}
\end{equation}
where $\vec{\hat z}$ is the unit vector in the direction of the rotation 
axis and $\vec{\hat x}$ is a unit vector in the equatorial plane. 
The wave number of the perturbation is $k=4\pi$. We added $\pi/4$ 
to the second term in (\ref{perturbation}) in order to provide mixed 
parity to the system. Equatorial and axial symmetry are thus broken
allowing the system to develop flows and fields in all modes. 

The initial magnetic field contains $B_0=100$ and $\epsilon = 1$ in 
most of our simulations. At ${\rm Rm}=10\,000$, this configuration 
implies a magnetic energy which is two orders of magnitude smaller than 
the kinetic energy in the initial rotation (precisely $E_{\rm mag}=
0.017 E_{\rm kin}$). The magnetic energy is thus much smaller
than the rotational energy as required for the onset of the
magneto-rotational instability. In a real stellar environment, these
quantities are roughly 30~orders of magnitudes apart.

We have also reduced the perturbation amplitude to $\epsilon=0.1$
in order to estimate the sensitivity of the results to this parameter.
As we will see in Section~4, the onset of the magneto-rotational
instability is not affected by the smaller amplitude. The decay 
of differential rotation takes slightly longer with a factor of 
1.6~but does not increase by an order of magnitude. 

Are stars in the efficient regime of the magneto-rotational
instability? For accretion disks rotating according to the
Keplerian law where $\Omega\sim s^{-3/2}$, the wavelength
of the most unstable mode is 
\begin{equation}
\lambda_{\rm inst} = 2\pi\sqrt{16/15}\,u_{\rm A}/\Omega
\label{lambda}
\end{equation}
where $u_{\rm A}=B/\sqrt{\mu\rho}$ is the Alfv\'en velocity 
(Balbus \& Hawley 1998). Let us assume that the rotation
profile is partly Keplerian in a star and the rotation
period be 1~day. A magnetic field of 100~kG will result --
for densities between 1 and $10^2$~g/cm$^{3}$ -- in wavelengths
of 10~to 100~km. 

At this point, we can see that the magneto-rotational instability will 
set in much less promptly if $\lambda_{\rm inst}>R$, thus imposing an
upper limit to the magnetic field. In principle, Eq.~(\ref{lambda})
reflects the aforementioned fact that the magnetic energy must
be significantly smaller than the kinetic energy in order to
be called a ``weak field''.  For stars, this is no problem as
$\lambda_{\rm inst}$ and $B$ are proportional, and the limiting
magnetic fields for which $\lambda_{\rm inst}>R$ are many orders
of magnitudes larger than the fields assumed in the stellar interior.

There is no lower limit for $B$ in ideal MHD, but the non-vanishing
diffusivity $\eta$ leads to a minimum magnetic field necessary
for the magneto-rotational instability. Essentially, the growth
rate of a perturbation with $\lambda_{\rm inst}$ must be larger
than the decay rate of a structure of the same wavelength.
The former is independent of $\lambda$, whereas the latter
changes with $\lambda^{-2}$. Balbus \& Hawley (1998) give estimates 
of the minimum field for the limits of ${\rm Pm}\gg1$ and 
${\rm Pm}\ll1$. Both cases lead to $B\approx 1$~G in stellar 
interiors, assuming $\eta=1000$~cm$^2$/s.

Even though magnetic fields in Ap stars are well in the suitable
range for instability, we have to check the applicability of our
initial $B$ for the numerical model which requires a much larger 
magnetic diffusion. This strong diffusion makes the instability
window much narrower than it is in reality. Also the stellar 
$\lambda_{\rm inst}$ is invisible in the simulations due to limited 
resolution. 

Using the angular velocity of about 6000 at $s=0.6$, the lower, 
diffusive limit is $B=75$, while the upper constraint from the 
size $R$ of the domain is $B=920$. In the computational setup 
with ${\rm Rm}=10\,000$ and $B_0=100$, we obtain a wavelength 
of the most unstable mode of $\lambda_{\rm inst} = 0.1$ near
$s=0.6$. Under real conditions, the range of suitable fields 
spans many orders of magnitudes, since $R$ is huge and $\eta$ 
is extremely small.

The values of ${\rm Pm}$, ${\rm Rm}$, the initial magnetic field
strength $B_0$ and the amplitude of the perturbation, $\epsilon$,
are the free parameters in the equations. Stellar gases possess 
magnetic Prandtl numbers of ${\rm Pm}\approx0.01$.
Values different from unity are typically difficult to achieve
by numerical schemes. Values vastly different from unity mean
that the time-scales for the diffusive processes in velocity
and magnetic fields differ very much and are thus hard to cover
appropriately by one simulation. We will vary Pm in our MHD simulations
to evaluate the reasonableness of using Pm near unity.

The velocity and magnetic field in our computational domain are
decomposed into 50 Chebyshev polynomials, 100 Legendre polynomials,
and 30 Fourier modes. This resolution is sufficient to resolve
the above wavelength of $\lambda_{\rm inst} = 0.1$ in our 
numerical setup. The nonlinear terms $-(\vec u\cdot\nabla)\vec u 
+ (\nabla\times \vec B) \times \vec B$ and
$\nabla\times(\vec u\times \vec B)$ are computed on a suitable
number of collocation points in real space,
and the spectral decomposition of these ``right sides'' are
fed into the implicit time-stepping scheme of the
linear part of Eqs.~(\ref{ns}) and (\ref{induction}).
First simulations with lower resolutions of $(K,L,M)=(25,30,30)$ 
and $(K,L,M)=(50,50,30)$ led to the same results within 3\%.

The boundary condition for the flow is stress-free at the
innermost and outermost radius. Vacuum conditions are imposed 
to the magnetic field at the inner and outer boundaries.

\begin{figure}
\epsfig{file=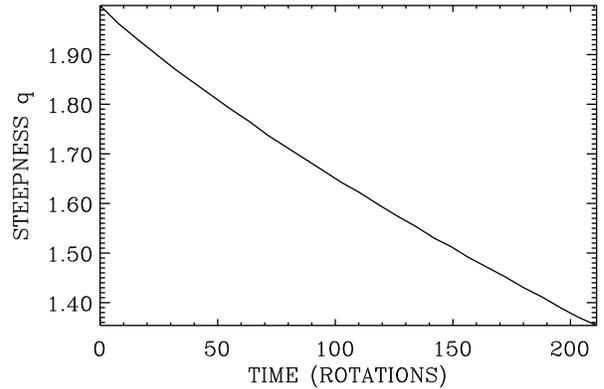,width=\linewidth}
 \caption{Hydrodynamical computation of the viscous decay 
 of differential rotation. A Reynolds number of ${\rm Rm}=50\,000$
was used here}
\label{spomega_q_g0}
\end{figure}

\section{\label{hydro}Hydrodynamic evolution}
Before studying the magnetohydrodynamic case, we have to assess
the evolution of the rotation flow without magnetic fields. This
is of particular interest since the stress-free boundary conditions 
are not compatible with the initial azimuthal velocity profile 
$u_\phi(r\sin\theta)$. The rotation profile will lead to meridional
circulations which equalize the differential rotation on the
viscous time-scale. We thus determine the purely hydrodynamic decay 
time before we can turn to magnetic configurations and their instabilities.

We measure a decay time with $\Omega(r,\phi)$ in the equatorial plane. 
The quantity is averaged over $\phi$, providing a one-dimensional
function of $r$. The equatorial plane was chosen just for simplicity.
Function~(\ref{omega}) is fitted to that profile
varying Rm and $q$. We determine the time when $q(t)=1$ and
call it the decay time $\tau_{\rm decay}$.

\begin{figure*}
\epsfig{file=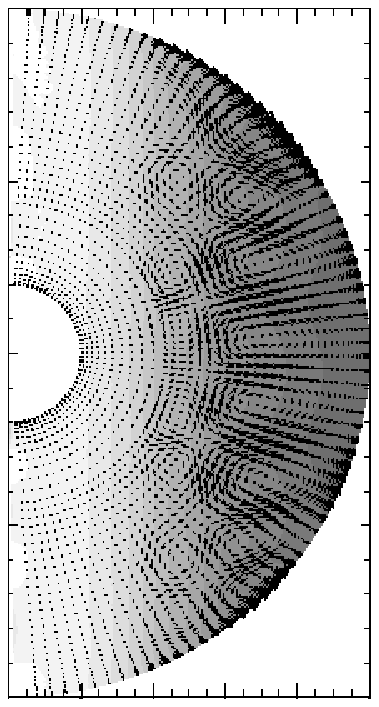,width=4.40cm}
\epsfig{file=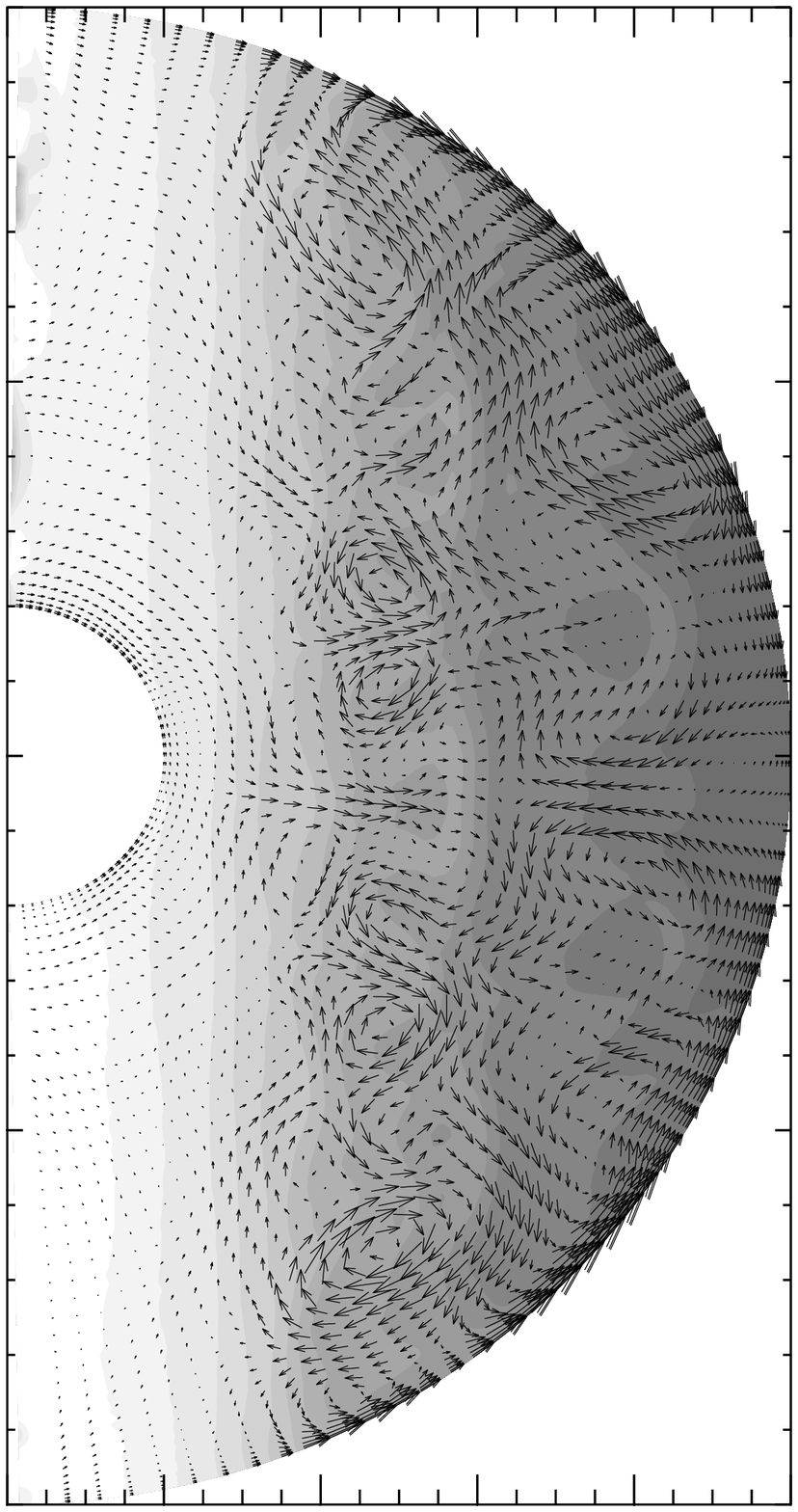,width=4.40cm}
\epsfig{file=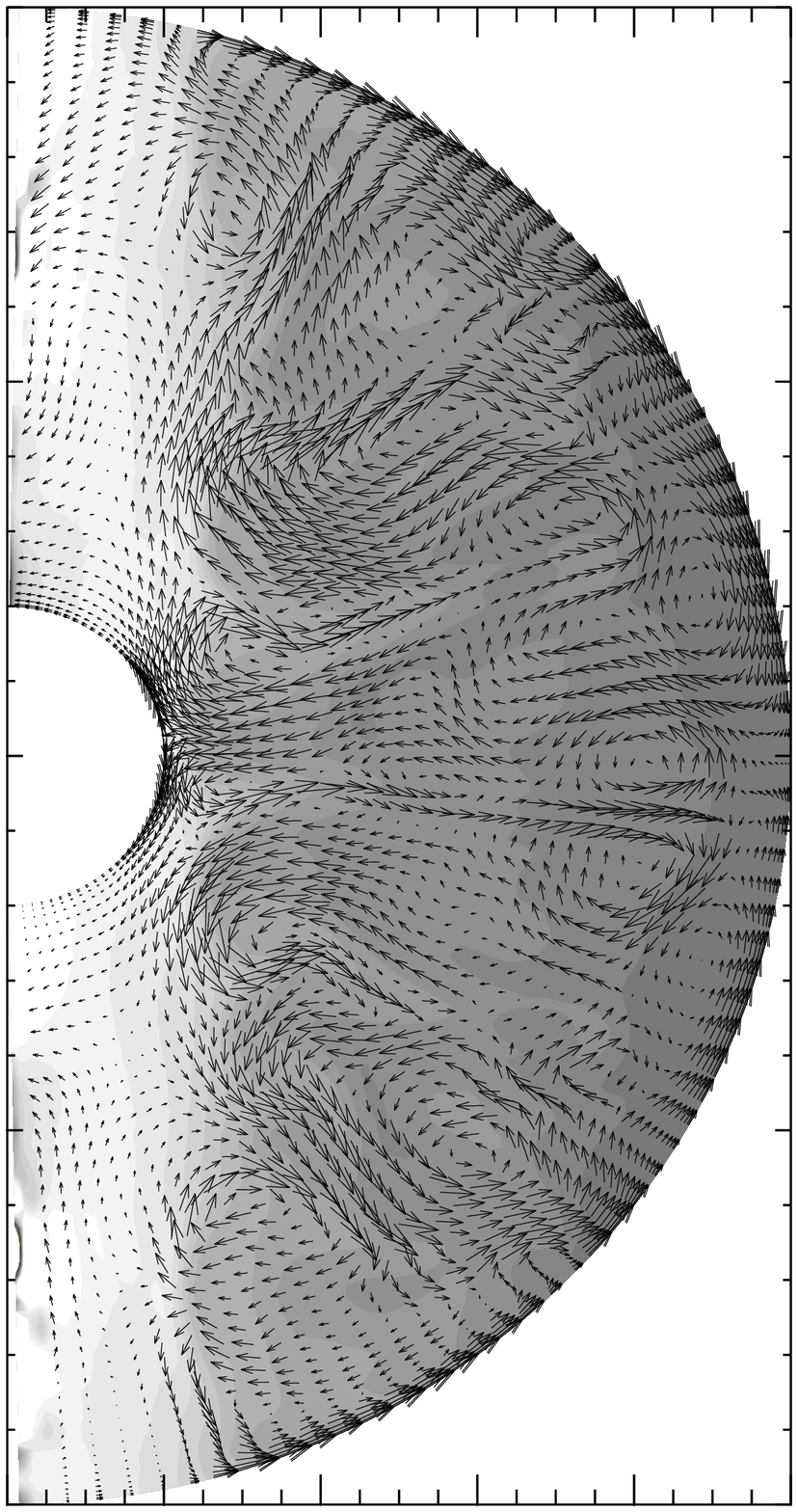,width=4.40cm}
\epsfig{file=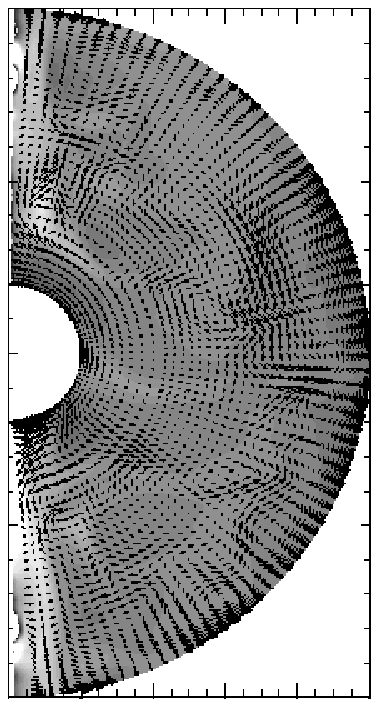,width=4.40cm}
\caption{Series of velocity snap-shots for the model with 
${\rm Rm}=10\,000$ and ${\rm Pm}=1$ 
at $t=0.001$, 0.002, 0.003, 0.004 diffusion
times (1.6, 3.2, 4.8, and 6.4~$\tau_{\rm rot}$). 
The grey level contours refer to the angular velocity $u_\phi/s$. 
The scaling of these contours is the same in all four 
cross-sections, while the scaling of the vector lengths 
varies. For better visibility, we plotted only
every second vector; only every third below $r=0.45$}
\label{series}
\end{figure*}

\begin{figure*}
\epsfig{file=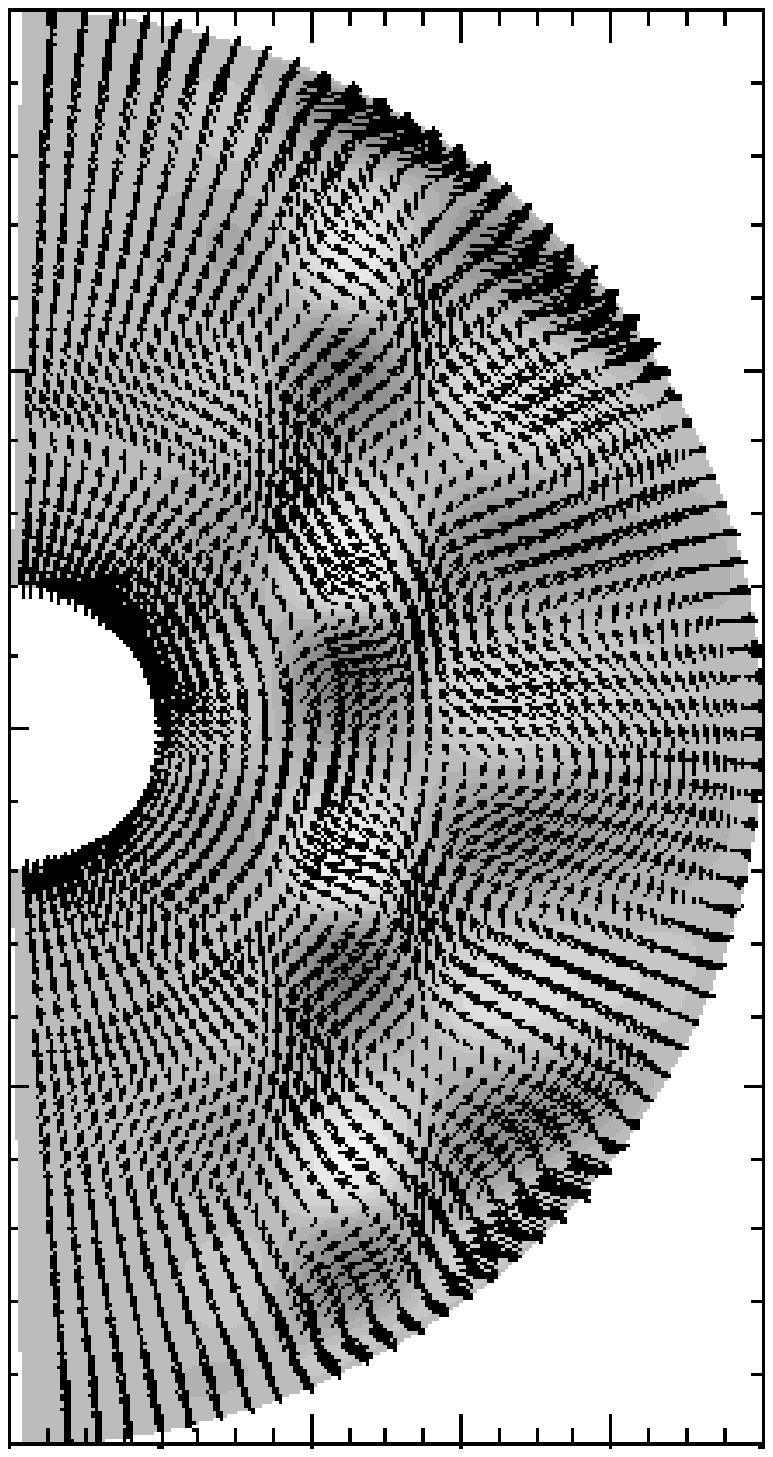,width=4.40cm}
\epsfig{file=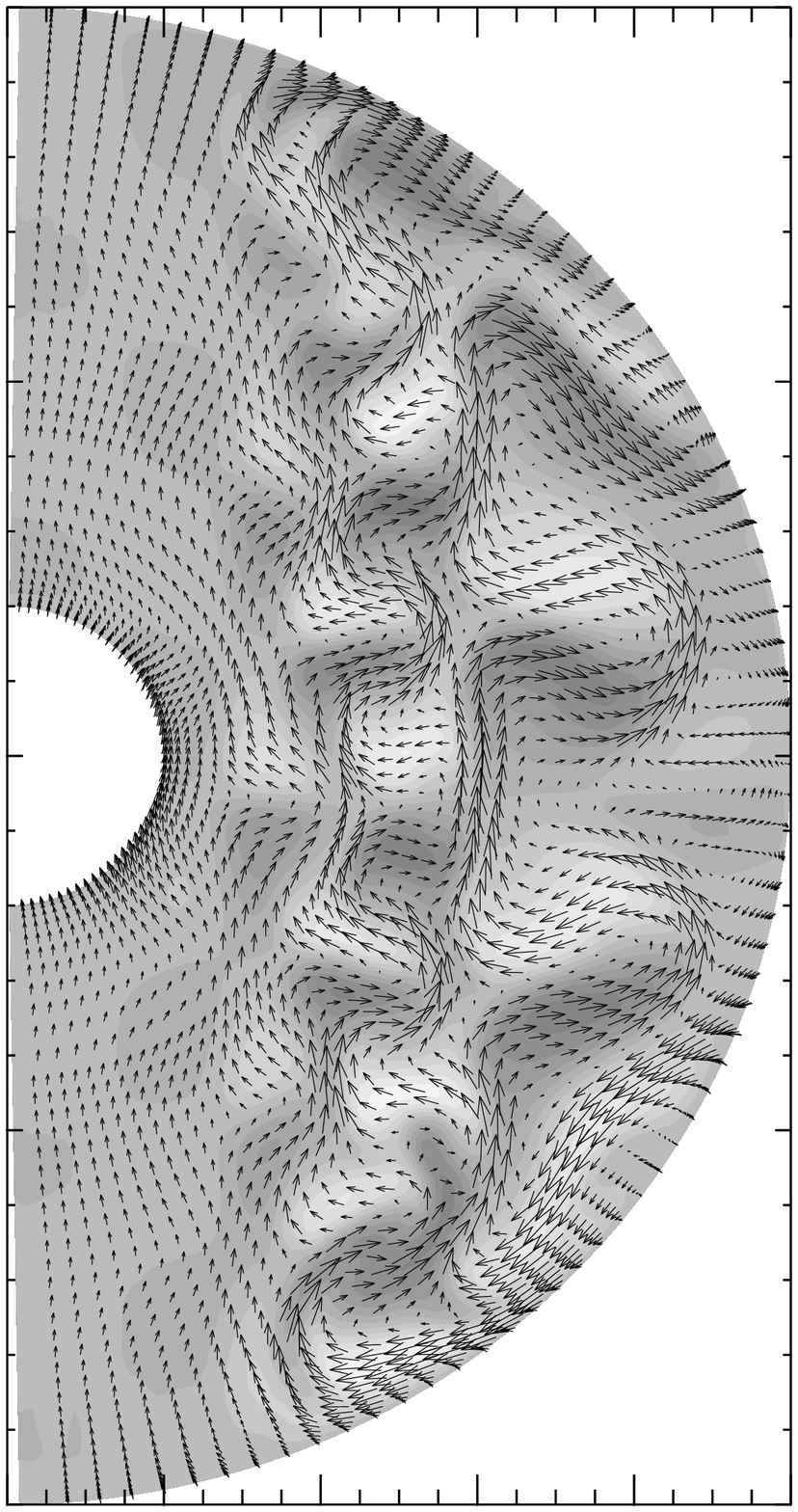,width=4.40cm}
\epsfig{file=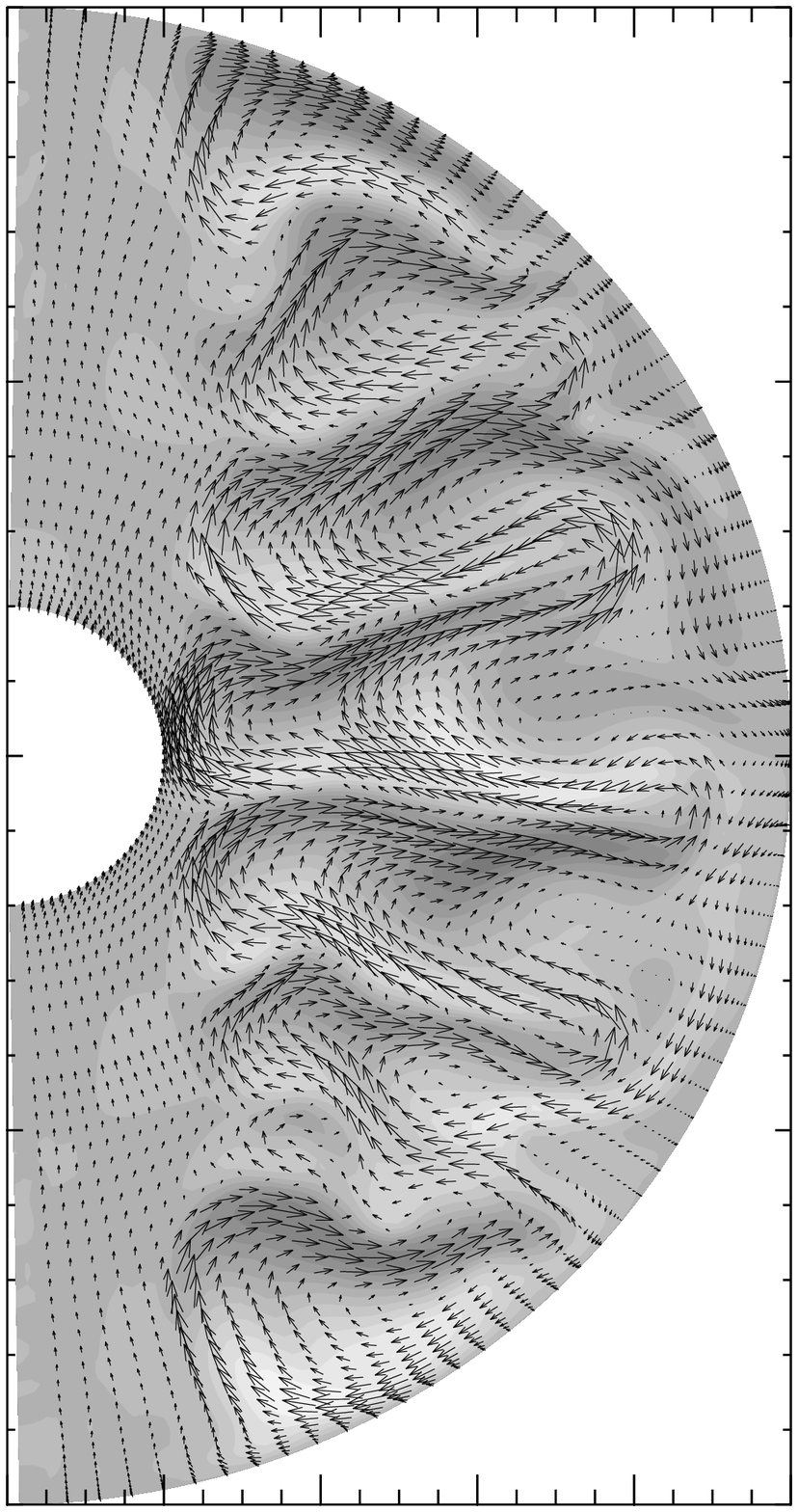,width=4.40cm}
\epsfig{file=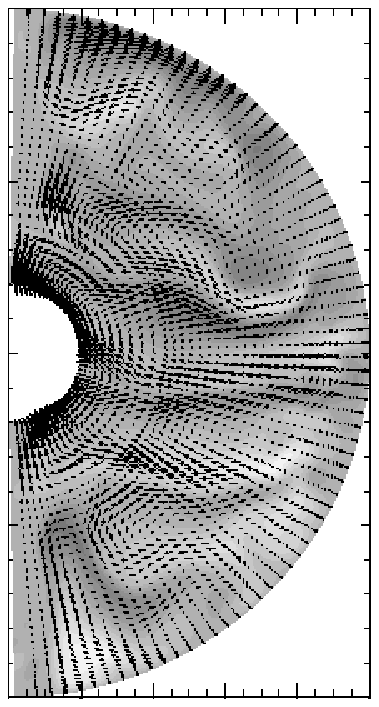,width=4.40cm}
\caption{Series of snap-shots of the magnetic field for the model with 
${\rm Rm}=10\,000$ and ${\rm Pm}=1$ 
at $t=0.001$, 0.002, 0.003, 0.004 diffusion
times (1.6, 3.2, 4.8, and 6.4~$\tau_{\rm rot}$). 
The grey level contours refer to $B_\phi$. 
Again we plotted only every second vector; 
only every third below $r=0.45$}
\label{seriesb}
\end{figure*}

A model with ${\rm Rm}=10\,000$ gradually decays and reaches $q=1$ 
after 65~rotation periods $\tau_{\rm rot}$. We will later see that this
time is much longer than the decay time caused by the magneto-rotational
instability. The kinetic energy in the meridional circulation is less than
1/2000 of the energy of the azimuthal velocities. At ${\rm Rm}=50\,000$ 
the viscous decay lasts much longer than 200~rotation periods 
(Fig.~\ref{spomega_q_g0}). The ratio of azimuthal to meridional 
kinetic energies now exceeds $10^4$. Viscosities as low as stellar 
microscopic values are of course not achievable by numerical simulations. 
We can turn to magnetohydrodynamic computations knowing that the
desired mechanism of differential-rotation decay should work on a 
time-scale of order 10 rotational periods or less to be essentially 
unaffected by viscosity.

\begin{figure}
\begin{center}
\epsfig{file=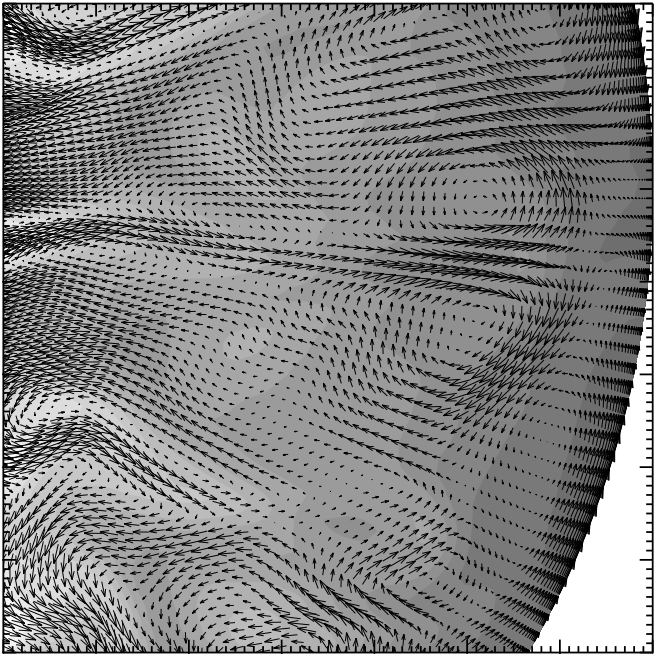,width=0.98\linewidth}
\end{center}
\caption{Detail of the velocity field shown in the third 
slice in Fig.~\ref{series} with full collocation point 
resolution}
\label{detail}
\end{figure}

\section{MHD evolution}
\subsection{Simulation results}
The following simulations including magnetic fields regard the 
enhanced decay rate of differential rotation as expected from the 
magneto-rotational instability. An initial magnetic field as 
described in Eq.~(\ref{perturbation}) in Section~\ref{setup}
causes poloidal flows with vortices of the same size as the
perturbation of the magnetic field. They form in all places,
except where the gradient of the angular velocity is negligible,
i.e.\ close to the rotation axis. A series of vertical cuts
through the velocity field at four equidistant times is
shown in Fig.~\ref{series}. The projected velocity vectors
indicate that the problem is numerically resolved. Note
that only every second vector of our collocation point grid
is plotted for the sake of visibility; at radii smaller than 
0.45 only every third vector. The same series is shown for
the magnetic field in Fig.~\ref{seriesb}. The vector lengths 
in both graphs are not comparable among the four slices; they 
are scaled for best visibility.

\begin{figure}
\epsfig{file=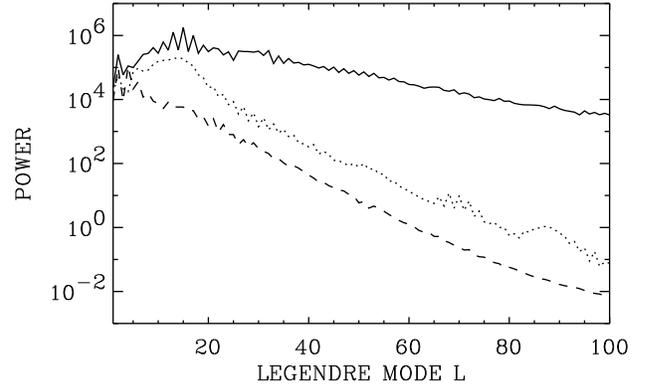,width=\linewidth}
\caption{Legendre spectra (latitudinal decomposition) of the 
energy of the magnetic field for the model 
with ${\rm Rm}=10\,000$ and ${\rm Pm}=1$ at three instances.
Dotted line: $t=0.001$ in the beginning of the instability;
solid line: $t=0.003$ during the period of strongest flows
(transition period); 
and the dashed line: $t=0.010$ when velocity and magnetic
fields decay. These times correspond to 1.6, 4.8, and 
$16~\tau_{\rm rot}$}
\label{lspectrum}
\end{figure}

The first velocity snap-shot after 1.6~orbital revolutions shows
the development of cells as the direct consequence of the
Lorentz forces resulting from the magnetic-field perturbation.
Two counter-rotating vortices represent one wave length of the 
perturbation from the second term 
in~(\ref{perturbation}). Roughly four of these waves 
fit into the sphere according to the wavenumber $k=4\pi$. The second 
slice shows the emergence of relatively thin sheets of strong radial
flows in up and down directions. These features become very prominent
in the third figure after $4.8~\tau_{\rm rot}$. They are
actually quite extended over several tens of degrees in 
azimuthal direction. A detailed plot with the full resolution
of our collocation point grid is shown in Fig.~\ref{detail}
magnifying a localized upstream. The fourth velocity slice 
of Fig.~\ref{series} shows an almost equalized rotation 
profile and a decay of small-scale features in the flow.

\begin{figure}
\epsfig{file=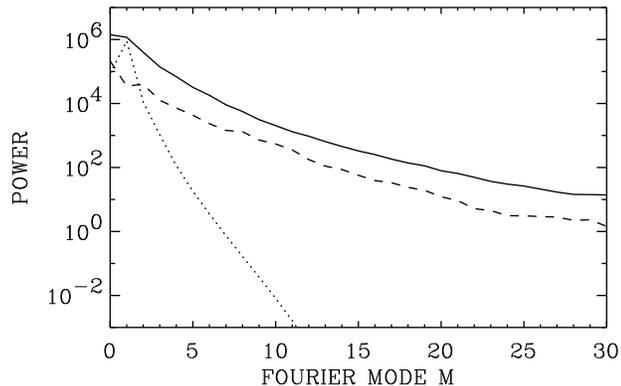,width=\linewidth}
\caption{Azimuthal decomposition in terms of Fourier spectra 
of the magnetic field energy for the model with ${\rm Rm}=10\,000$ 
and ${\rm Pm}=1$ at the same times $t=0.001$ (dotted), 0.003 
(solid), and 0.010 (dashed) as in Fig.~\ref{lspectrum}}
\label{mspectrum}
\end{figure}

The latitudinal resolution of the model shown in Fig.~\ref{seriesb}
is plotted in Fig.~\ref{lspectrum} as a series of three 
Legendre spectra at $t=0.001$, 0.003,
and 0.010, corresponding to $1.6~\tau_{\rm rot}$, $4.8~\tau_{\rm rot}$,
and $16~\tau_{\rm rot}$ resp. Maximum and minimum power
span 2.5 orders of magnitudes in the most turbulent case (solid line).
The initial power contrast is $10^7$ (dotted line); after the redistribution
of angular momentum, the contrast quickly reaches the
same order of magnitude again (dashed line in Fig.~\ref{lspectrum}).
The spectra of the velocity fields are very similar.
The Fourier spectra as given in Fig.~\ref{mspectrum}
show very satisfying power contrast all through the
simulation.

\begin{figure}
\epsfig{file=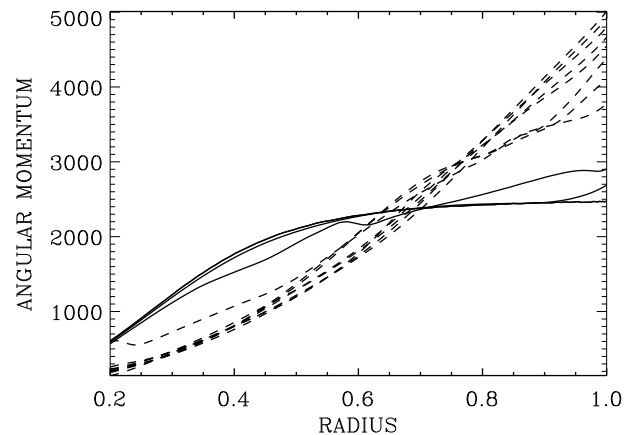,width=\linewidth}
 \caption{Redistribution of angular momentum $j$ in the model with
 ${\rm Rm}=10\,000$ and ${\rm Pm}=1$. The same equatorial averages as 
 applied for the determination of the steepness $q$ were used, and 
 $j$ was plotted for various times of the simulation in steps of 
 $1.6\tau_{\rm rot}$; dashed lines are for $t\geq 4.8\tau_{\rm rot}$. 
 The flat profile is from the initial rotation 
 profile, the steep, nearly parabolic profile is the final state with 
 almost constant angular velocity}
\label{spmom_time}
\end{figure}

The change of specific angular momentum as a function
of axis distance is plotted in Fig.~\ref{spmom_time}.
The profiles of $s^2\Omega$ are -- for simplicity -- 
again taken from the equatorial plane and averaged over
the $\phi$-direction. The initial differential rotation
profile of (\ref{omega}) with ${\rm Rm}=10\,000$ and
$q=2$ is shown as a flat curve, while the steepest,
nearly parabolic lines are the final distribution of
angular momentum, corresponding to a nearly uniform rotation.
We plotted dashed lines for $t\geq4.8\tau_{\rm rot}$;
a gap between these and the solid lines marks the transition
period when strongest transport of angular momentum is
found. 

The redistribution of angular momentum is a combined result
of stresses from velocity and magnetic field fluctuations.
The averages $\langle u_r' u_\phi'\rangle$ and 
$-\langle B_r' B_\phi'\rangle$, which are again taken in the 
equatorial plane only, show a clear domination of magnetic
stresses over kinetic stresses, occasionally by a factor 10.


\begin{figure}
\epsfig{file=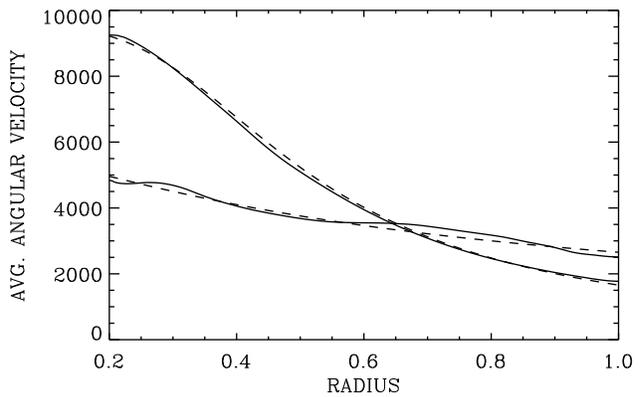,width=\linewidth}
\caption{Two examples of the angular velocity averaged
in the equatorial plane as a function of radius (solid
lines) and the fit of (\ref{omega}) with ${\rm Rm}$ and $q$ 
(dashed lines) at $t=0.0015$ ($2.4\tau_{\rm rot}$)
and $t=0.0032$ ($5.1\tau_{\rm rot}$)}
\label{spomega_q_plotfit}
\end{figure}

In the same way as in Section~\ref{hydro} we measure
the decay time of the differential rotation by the time
it takes the system to cross a $q=1$ profile. Despite
the enormous flows emerging, the azimuthally averaged
angular velocity provides us with profiles $\Omega(r)$
in the equatorial plane which are fairly compatible with
the two-parameter function~(\ref{omega}). 
Fig.~\ref{spomega_q_plotfit} shows examples of such fits
for two velocity snap-shots of the simulation illustrated
by snap-shots in Figs.~\ref{series} and~\ref{seriesb}. It is 
thus still reasonable to use the steepness $q$ for defining the decay
time even for the simulations where magneto-rotational
instability emerges.

\begin{figure}
\epsfig{file=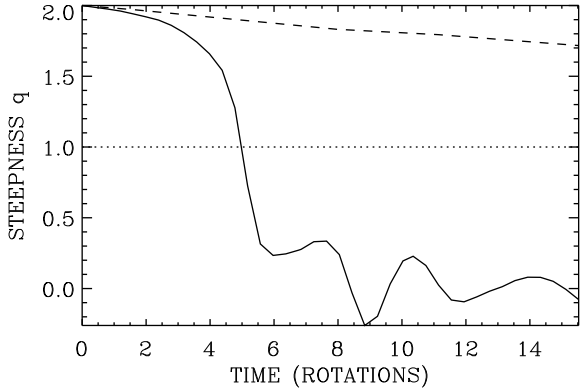,width=\linewidth}
\caption{Variation of the steepness of the rotation profile
with time as computed from azimuthal averages of the angular 
velocity of the model with ${\rm Rm}=10\,000$, ${\rm Pm}=1$, 
$B_0=100$, and $\epsilon=1$. The resulting function of $r$ was 
fitted to a function of the form (\ref{omega}). The dotted line
indicates $q=1$ which will be used to define the decay time of
the differential rotation. The dashed line shows the change of
$q$ in a hydrodynamic model where the differential rotation
decays only by viscosity}
\label{spomega_q}
\end{figure}

The decay of $q$ with time is given by the solid line in 
Fig.~\ref{spomega_q}. A short transition
period between $4.0~\tau_{\rm rot}$ and $5.5~\tau_{\rm rot}$
can be seen. The third velocity slice of Fig.~\ref{series}
with strong radial up and down flow sheets falls right in the 
middle of this period. Towards the end of the computation,
the system oscillates around an equilibrium state with $q=0$,
and magnetic and kinetic energies decay exponentially.

The transition period is less marked
when we go to lower diffusivities, i.e.\ to higher magnetic
Reynolds numbers Rm. The steepness $q$ diminishes more
gradually, but still on a time-scale which is an order of
magnitude shorter than viscous decay in purely hydrodynamic
simulations. (Note, however, that the difference to viscous
decay is expected to be much larger in a real star.) We have
added the non-magnetic model with ${\rm Rm}=10\,000$ as a 
dashed line in Fig.~\ref{spomega_q} illustrating the marginal 
influence of viscous decay on the MHD simulations.

\subsection{Application to stellar parameters}
\begin{figure}
\epsfig{file=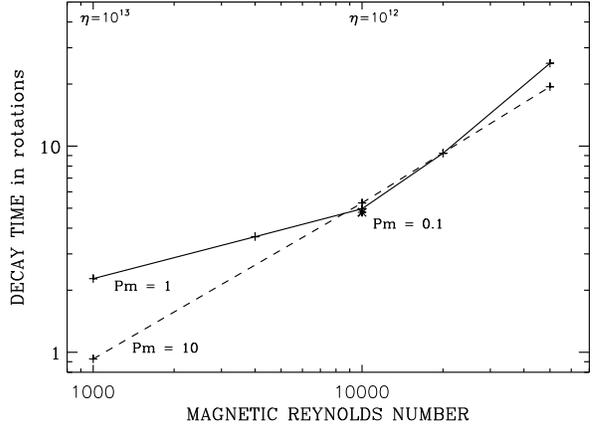,width=\linewidth}
 \caption{Decay time of differential rotation versus magnetic
 Reynolds number for nine simulations with high spectral 
 resolution. The decay time is measured in rotations. The
 upper abscissa gives an idea of the magnetic diffusivities $\eta$
 implied by the Reynolds numbers, assuming a stellar radius and
 stellar rotation rate. The solid line refers to ${\rm Pm}=1$, 
 the dashed line shows the results of ${\rm Pm}=10$. An asterisk 
 indicates the decay time for the only computation with ${\rm Pm}=0.1$}
\label{apdec_rm}
\end{figure}

The rotation profile seems to decay on the rotational time-scale. 
This is apparently way too fast for any trace of $\Omega(s)$-rotation 
in stars with radiative envelopes. We will later see that this 
is not quite true. There are the real physical quantities
which are hard to match in a computer simulation. The diffusive
time-scale is orders of magnitudes longer than the rotational
time-scale. We achieved to make them more than four orders
of magnitudes different and may obtain an extrapolation
towards real stellar parameters. The magnetic Reynolds number
${\rm Rm} = {R^2\Omega}/{\eta}$
in stellar radiative zones is about $10^{13}$--$10^{14}$. 
The highest Rm achieved numerically is 50\,000 in this presentation.
The dependence of the decay time $\tau_{\rm decay}$ on the magnetic 
Reynolds number and magnetic Prandtl number is shown in 
Fig.~\ref{apdec_rm}. The decay times are given in rotation periods 
which is a few days for CP stars. Fortunately, we found no 
significant dependence of the decay times on Pm. As the true
magnetic Prandtl number will be of the order of $10^{-2}$ for
stars, we may assume that our Pm near unity will not imply
severe differences from the real physics.

\begin{figure}
\epsfig{file=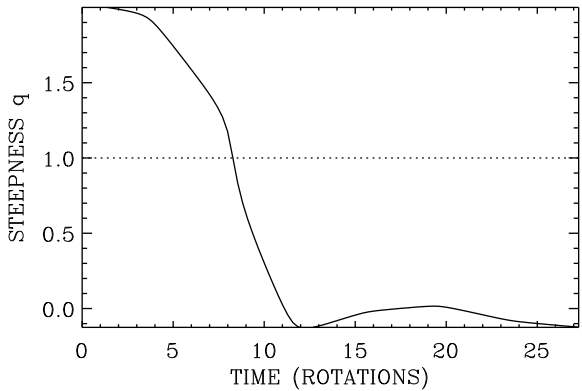,width=\linewidth}
\caption{Variation of differential rotation of the model with 
${\rm Rm}=10\,000$ as in Fig.~\ref{spomega_q}. The initial
perturbation of the vertical magnetic field, however, was a 
factor 10 smaller than there, $\epsilon=0.1$}\label{spomega_q_lowb}
\end{figure}

Also the amplitude $\epsilon$ of the perturbation of the
initial vertical magnetic field was changed to $\epsilon=0.1$.
The resulting decay of differential rotation for the model
with ${\rm Rm}=10\,000$ and ${\rm Pm}=1$ is shown in 
Fig.~\ref{spomega_q_lowb} in terms of the steepness $q$
of the rotation profile. The decay time increased from 5~to
8~rotation periods. Also the ratio of magnetic to kinetic
stresses as the constituents of the transport of angular
momentum is not changed and reaches 10:1.

With a series of computations with various Rm, we can make an
attempt in extrapolating the decay time to stellar conditions.
It is found that the decay time scales as the magnetic Reynolds
number, in particular for ${\rm Rm}\geq 10\,000$ which is the 
interesting interval. Measured in diffusion times, the decay 
time depends only slightly on Rm. We derive  the
relation
\begin{equation}
  \frac{\tau_{\rm decay}}{\tau_{\rm rot}} = \frac{{\rm Rm}}{2000}
  \label{rm_relation}
\end{equation}
where $\tau_{\rm rot}$ is the rotation period of the star. If the 
stellar parameters $R=2R_\odot$, $\tau_{\rm rot}=1$~day, and 
$\eta=1000$~cm$^2$/s (Spitzer 1956) are applied, the magnetic Reynolds
number is $3\cdot10^{14}$. The diffusivity is the variable
ingredient here; our diagram in Fig.~\ref{apdec_rm} can thus
be annotated with $\eta$ in stead of Rm (see upper abscissa).
If Rm enters (\ref{rm_relation}) with an exponent unity, the 
decay time is actually locked to the diffusion time and not 
to the rotation period. If we adopt a stellar ${\rm Pm}=10^{-2}$
and rewrite Eq.~\ref{rm_relation} as
${\tau_{\rm decay}}/{\tau_{\rm rot}} = 0.0005\,{\rm Pm}\,{\rm Re}$,
we see that the differential-rotation decay as a consequence
of the magneto-rotational instability is of $O(10^5)$ times
faster than the viscous decay scaling with Re.

Relation~(\ref{rm_relation}) delivers a decay time of
$3\cdot10^8$~yr for the stellar parameters given above.
This is of the order of the life-time of an A star. This
extrapolation bears of course a wide uncertainty. If we
look at the graph for ${\rm Pm}=10$ the power of Rm is
slightly lower, $\tau_{\rm decay}/\tau_{\rm rot} = 
0.002{\rm Rm}^{0.85}$. The resulting decay time for ${\rm Rm}=3\cdot10^{14}$
is $\tau_{\rm decay} = 10^7$~yr.

\section{Summary}
Since differential rotation in radiative stellar zones cannot
be damped by viscosity in a life-time of a star, we investigate
the magnetic evolution which is likely to imply the onset of 
the magneto-rotational instability providing efficient 
angular-momentum transport. MHD simulations of spherical shells
were performed showing that the instability emerges indeed, and
quick equalization of differential rotation is found.
An extrapolation to stellar parameters gives decay times of 
differential rotation of the order of 10--100 million years.
This is the time-scale on which redistribution of angular momentum 
is taking place. The magneto-rotational instability grows on
a scale of rotation periods, but the nett efficiency of angular
momentum transport in the fully nonlinear regime is obviously
a different one.

While solar-type stars have long enough life-times to rotate 
uniformly in their radiative cores, the life-time of stars of 
spectral type A is of the same order of magnitude as the decay 
time of about 100 million years. CP stars have magnetic fields
providing the conditions for the magneto-rotational 
instability -- along with an initial differential rotation
caused by interactions with the accretion disk of the pre-main
sequence life. Because of the comparable stellar life-time and
decay time, differential rotation should thus be present in CP
stars during a considerable period of their life. As long as the total
angular momentum is conserved, we should expect a slight
{\it increase\/} of surface angular momentum with age. 
Since an A star roughly doubles its radius during the presence
on the main sequence, the resulting decrease of surface
rotation may balance with the increase due to the mechanism
described in this paper. Age data and precise rotation periods 
of CP stars will be needed to test this result.

\end{document}